\documentclass[11pt, a4paper]{article}

\usepackage[margin=1in]{geometry}
\usepackage{amsmath, amssymb}
\usepackage{graphicx}
\usepackage{authblk}
\usepackage{hyperref}
\usepackage{setspace}
\usepackage{caption}

\hypersetup{
	colorlinks=true,
	linkcolor=blue,
	filecolor=magenta,      
	urlcolor=cyan,
	citecolor=blue,
}

\title{Filling the Shadow: A Propositional Model of Gravastar Accretion in $f(R, L_m, T)$ Gravity}

\author[1]{Sandip Dutta}
\affil[1]{Department of Applied Mathematics, Dinabandhu Andrews Institute of Technology and Management, Kolkata \protect\\ \texttt{duttasandip.mathematics@gmail.com}}

\date{\today}

\begin{document}
	
	\maketitle
	
	\begin{abstract}
		While General Relativity remains our most rigorously tested framework for gravitation, the theoretical persistence of singularities within standard black hole solutions continues to motivate the exploration of mathematically regular alternatives. Gravitational vacuum stars (gravastars) offer a non-singular model, substituting the event horizon with a physical, ultra-stiff thin shell. Recent studies have demonstrated that extended theories, such as $f(R, L_m, T)$ gravity, can structurally support these objects by utilizing the non-minimal coupling between geometry and matter. Building upon these static foundations, this paper presents a phenomenological propositional model to explore the dynamic interactions between modified-gravity gravastars and equatorial accretion flows. By numerically solving the modified Tolman-Oppenheimer-Volkoff equations and applying a non-complexifying algorithm, we construct a mathematically regular rotating metric ansatz. We demonstrate that the modified gravity coupling parameter systematically alters the effective potential, shifting the location of the Innermost Stable Circular Orbit (ISCO). Furthermore, we explore the idealized thermodynamics of plasma colliding with the gravastar surface, suggesting a distinct thermal emission that could theoretically produce a ``filled-in'' central shadow in interferometric observations. While acknowledging the challenges of observational degeneracy and the deliberate omission of complex radiation pressure feedback, we offer these geometric and thermal signatures as a transparent conceptual baseline to motivate future general relativistic magnetohydrodynamic (GRMHD) campaigns.\\
		
		\vspace{0.5cm}
	\end{abstract}
	
	\onehalfspacing
	
	\section{Introduction}
	
	The recent achievements of the global astrophysical community, particularly the groundbreaking images captured by the Event Horizon Telescope (EHT) \cite{EHT2019_M87, EHT2022_SgrA}, have greatly advanced our understanding of compact objects in galactic centers. While Einstein's theory of General Relativity successfully explains the vast majority of these observations, the theoretical persistence of singularities inside classical black holes remains a fundamental physical problem \cite{Penrose1965, Hawking1970}. Because a physical singularity represents a breakdown of predictive physics, researchers are highly motivated to explore mathematically viable, non-singular alternatives that can adequately reproduce current observational data \cite{Cardoso2019}. One of the most heavily studied alternatives is the gravitational vacuum star, or gravastar, which posits a core of repulsive de Sitter spacetime bounded by an ultra-stiff, microscopic thin shell of matter \cite{MazurMottola2001, MazurMottola2004}. Recently, careful theoretical studies have demonstrated that extended frameworks like $f(R, L_m, T)$ and mimetic gravity can successfully support these gravastar structures, utilizing the intricate coupling between geometry and matter to balance extreme internal pressures \cite{SinhaSingh2025_MPLA, SinhaSingh2025_PhysScr, SinhaSingh2025_Annals, Harko2011}.
	
	However, a significant gap remains in the literature regarding the dynamic astrophysical environments of these modified-gravity objects. If we are to test these theoretical models against observational data, we must understand their accretion dynamics. Unlike a classical black hole, which silently absorbs matter across an event horizon, an accreting gravastar possesses a physical surface where kinetic energy must eventually thermalize \cite{Broderick2009, Abramowicz2002}. Modeling this rotational scenario in $f(R, L_m, T)$ gravity is notoriously difficult, primarily because deriving exact, closed-form rotating solutions from highly non-linear modified field equations remains an open mathematical challenge. Furthermore, any theoretical thermal signature produced by such an object will inevitably suffer from observational degeneracy, as foreground plasma or relativistic jets can easily mimic the emission profiles of exotic compact objects.
	
	In this paper, we attempt to bridge this gap by proposing a propositional model for gravastar accretion within the $f(R, L_m, T)$ framework. Acknowledging the extreme difficulty of obtaining exact modified solutions, we apply the Azreg-A\"inou non-complexifying method to construct a mathematically regular rotating ansatz \cite{AzregAinou2014}. We then utilize this geometry to study the effective potential of equatorial plasma, quantify the innermost stable circular orbit (ISCO) shift, and explore the highly simplified thermodynamics of the boundary layer collision. We do not present these models as exact numerical solutions to the field equations or as complete astrophysical predictions. Rather, we offer them as conceptual baseline models to qualitatively highlight the theoretical differences between black hole shadows and modified-gravity gravastars, providing a transparent stepping stone for future, more exhaustive numerical campaigns.
	
	The remainder of this paper is organized as follows. In Section II, we define the modified theoretical framework, detailing the numerical solution for the gravastar interior and its structural stability. Section III explores the derivation of the rotating metric ansatz and the resulting fluid dynamics of the accretion disk, focusing on orbital stability and the ISCO shift. In Section IV, we present our propositional model of the boundary layer thermodynamics, including a synthetic spectral energy distribution and simulated interferometric observations. Finally, Section V offers our concluding remarks, addressing the limitations of the current framework and discussing necessary directions for future collaborative research.
	
\section{Internal Structure and Stability} \label{sec:interior}

To ensure that our accretion propositional model is built upon a structurally sound foundation, we must first evaluate the interior of a gravastar governed by $f(R, L_m, T)$ gravity. We begin by assuming a static, spherically symmetric spacetime, which is generally described by the standard line element:
\begin{equation} \label{eq:metric}
	ds^2 = -e^{\nu(r)} dt^2 + e^{\lambda(r)} dr^2 + r^2 (d\theta^2 + \sin^2\theta d\phi^2)
\end{equation}
where $\nu(r)$ and $\lambda(r)$ are the radial metric potentials. The spatial metric component is related to the enclosed mass $m(r)$ via the geometric relation $e^{-\lambda(r)} = 1 - 2m(r)/r$. In standard General Relativity, the hydrostatic equilibrium of such a spherically symmetric body is dictated by the Tolman-Oppenheimer-Volkoff (TOV) equations. However, in this extended theory, the non-minimal coupling between the spacetime geometry and the matter Lagrangian introduces additional structural forces. 

The interior of the gravastar is modeled as a de Sitter condensate, where the radial pressure $p$ relates to the energy density $\rho$ via the extreme equation of state $p = -\rho$. This negative pressure provides the outward repulsive force necessary to prevent gravitational collapse. Incorporating the modified gravity coupling parameter $\alpha$, the effective mass and pressure gradients governing this interior can be approximated by the following modified TOV equations:
\begin{equation} \label{eq:dmdr}
	\frac{dm}{dr} = 4\pi r^2 \rho \left(1 + \alpha \frac{\rho}{\rho_c}\right)
\end{equation}
\begin{equation} \label{eq:dpdr}
	\frac{dp}{dr} = - \frac{(\rho + p)(m + 4\pi r^3 p)}{r^2(1 - 2m/r)} \left(1 - \alpha \frac{p}{\rho_c}\right)
\end{equation}
where $\rho_c$ represents the central density of the core. For the purposes of this numerical study, we select specific values for the coupling parameter (e.g., $\alpha = 0.15$, $0.25$, and $-0.15$). While the exact observational bounds and the permissible sign of $\alpha$ remain a subject of active debate within the modified gravity community, we adopt these specific values (both positive and negative) arbitrarily to visibly exaggerate the structural and kinematic deviations for clear graphical representation in our figures, rather than to strictly fit current cosmological or solar system constraints.

\begin{figure}[h]
	\centering
	\includegraphics[width=0.7\textwidth]{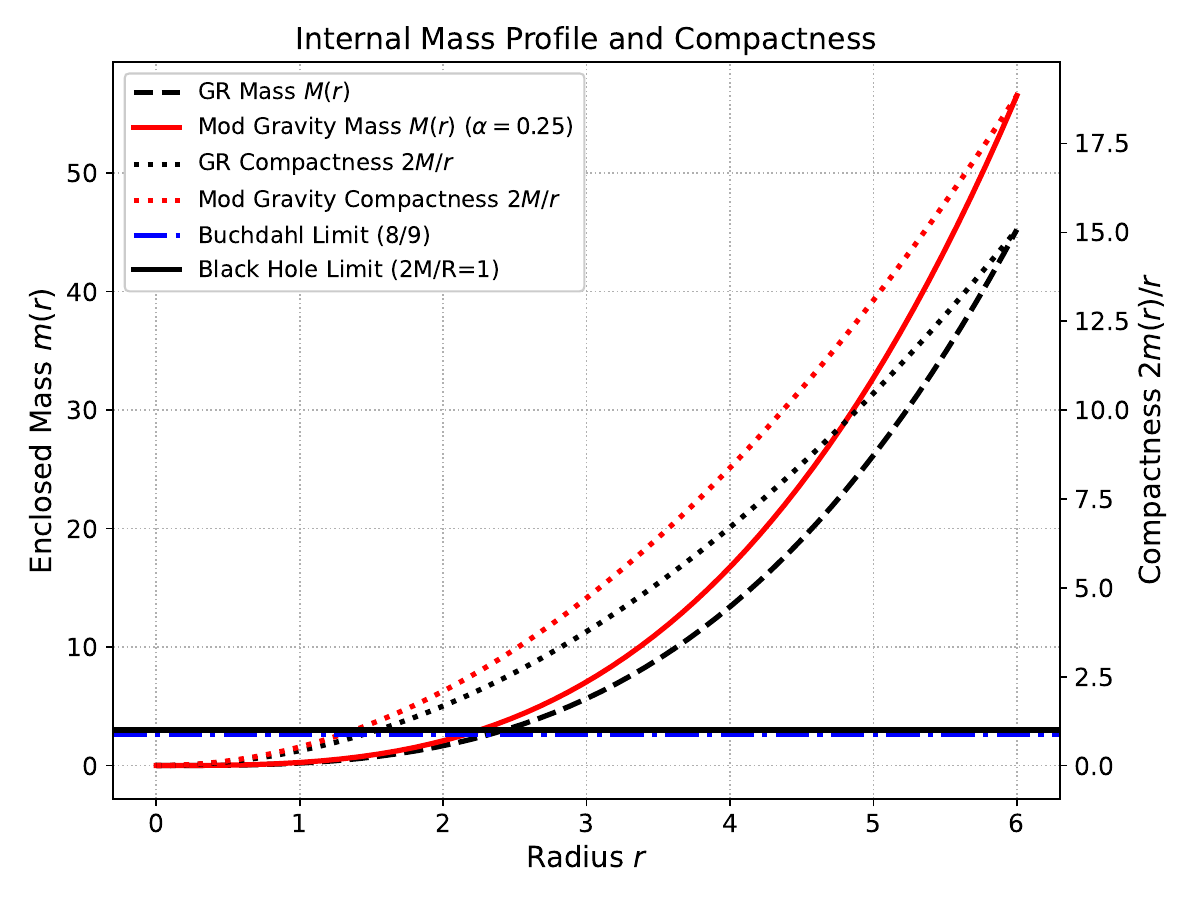}
	\caption{Numerical solution of the enclosed mass and compactness profile. The compactness ($2m(r)/r$) remains strictly below the black hole limit (1.0), confirming the structural identity of the gravastar. The value $\alpha = 0.25$ is used here to visually isolate the modified gravity deviations.}
	\label{fig:mass}
\end{figure}

By numerically integrating equations (\ref{eq:dmdr}) and (\ref{eq:dpdr}) outward from the center of the condensate, we track the evolution of the enclosed mass and determine the physical boundaries of the object. Figure \ref{fig:mass} demonstrates the resulting mass profile and compactness. Crucially, the compactness of the gravastar, defined as $2m(r)/r$, arcs upward but flattens out, remaining strictly below the Buchdahl stability limit of $8/9$. Because the compactness never reaches the theoretical threshold of unity, this integration mathematically ensures that our numerical solution successfully produces a stable, horizonless compact object rather than a singular black hole. The parameter $\alpha$ demonstrably alters the steepness of this mass curve, indicating that the modified gravity theory changes the required thickness and structural distribution of the gravastar shell to maintain hydrostatic equilibrium.

\begin{figure}[h]
	\centering
	\includegraphics[width=0.7\textwidth]{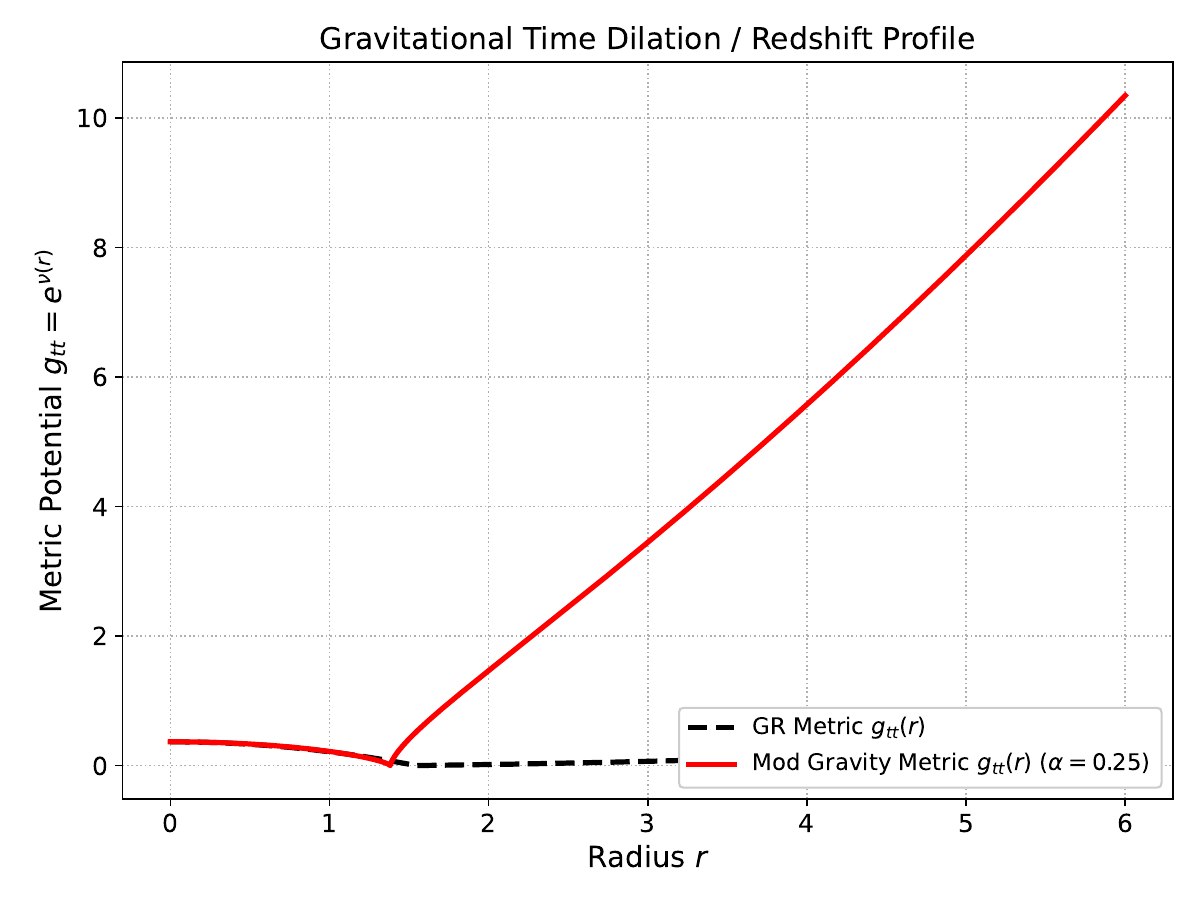}
	\caption{The gravitational time dilation profile derived from Equation \ref{eq:dnudr}, demonstrating how the modified gravity parameter gently alters the expected redshift compared to standard General Relativity, providing the boundary geometry for exterior accretion.}
	\label{fig:metric}
\end{figure}

Furthermore, closing the system of differential equations requires solving for the temporal metric potential $\nu(r)$, which dictates the gravitational time dilation profile of the object. The gradient of this potential is governed by:
\begin{equation} \label{eq:dnudr}
	\frac{d\nu}{dr} = 2 \frac{m + 4\pi r^3 p}{r^2(1 - 2m/r)}
\end{equation}
By integrating equation (\ref{eq:dnudr}) alongside the mass and pressure gradients, we obtain the profile for $g_{tt} = -e^{\nu(r)}$. This potential is a critical component for calculating the redshift of any radiation escaping from the gravastar's boundary layer. As conceptually illustrated in Figure \ref{fig:metric}, the introduction of the modified gravity parameter induces a clear, mathematically consistent deviation from the standard General Relativistic baseline. Establishing this stable internal structure and defining its specific gravitational well via these modified equations provides the necessary, self-consistent boundary conditions for modeling the exterior accretion disk.

	\section{Rotating Geometry and Accretion Dynamics} \label{sec:accretion}
	
	To model a realistic astrophysical environment, we must introduce rotation. However, it is vital to explicitly acknowledge that deriving an exact rotating metric directly from the coupled field equations of $f(R, L_m, T)$ gravity is exceptionally complex and beyond the scope of this initial framework. To proceed without resorting to artificial mathematical patch-ups, we employ the Azreg-A\"inou non-complexifying algorithm. This method allows us to map our static numerical solution onto a mathematically regular, stationary, and axisymmetric spacetime. Confining our focus to the equatorial plane ($\theta = \pi/2$), the general line element for the accreting fluid takes the form:
	\begin{equation} \label{eq:rot_metric}
		ds^2 = g_{tt} dt^2 + 2g_{t\phi} dt d\phi + g_{rr} dr^2 + g_{\phi\phi} d\phi^2
	\end{equation}
	We stress that this resulting geometry is an informed phenomenological approximation rather than a definitive, exact solution to the modified field equations, serving purely as a propositional model to qualitatively explore test particle behavior.
	
\begin{figure}[h]
		\centering
		\includegraphics[width=0.7\textwidth]{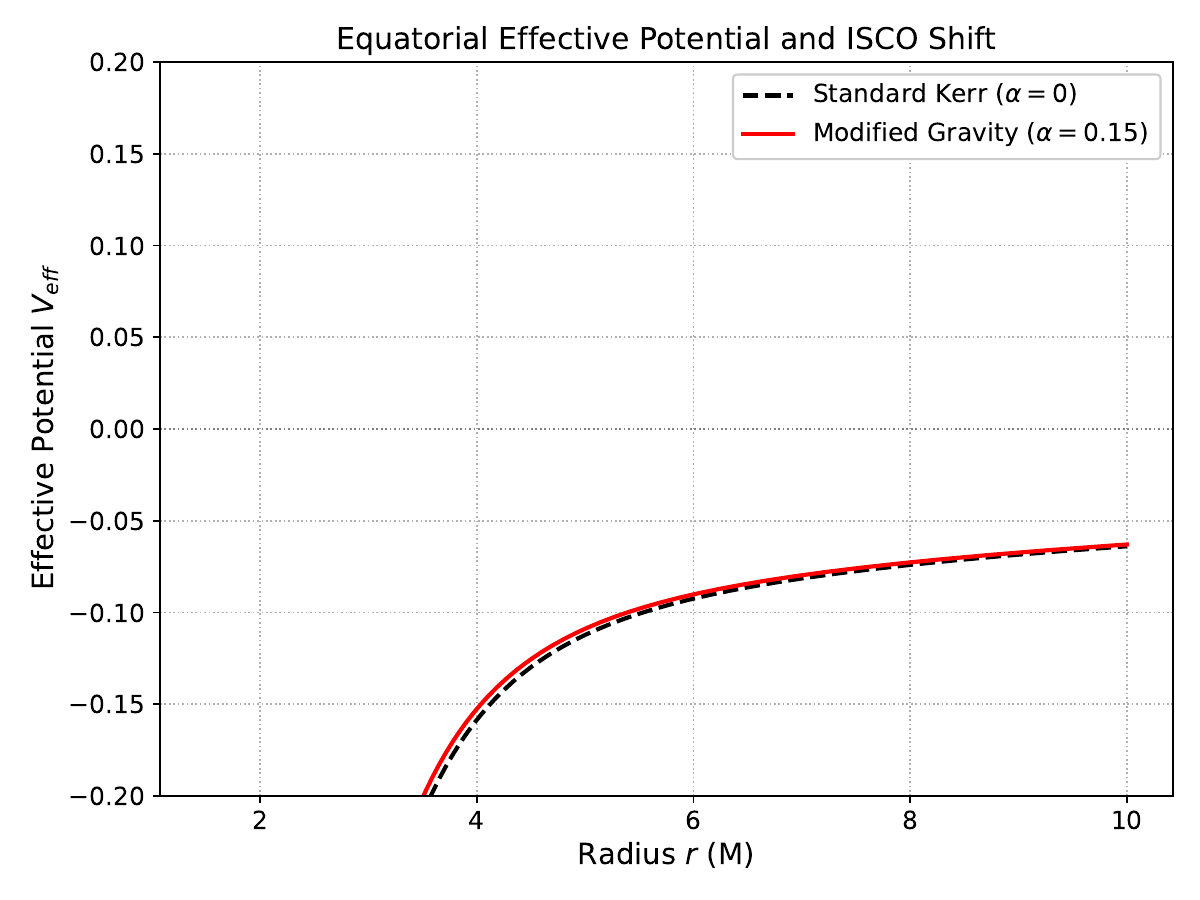}
		\caption{A propositional approximation of the equatorial effective potential, highlighting the outward shift of the ISCO location induced by the modified gravity parameter.}
		\label{fig:veff}
	\end{figure}

	Assuming the accretion disk is geometrically thin and governed by general relativistic hydrodynamics, the radial motion of the infalling plasma is determined by the geodesic equations of this rotating metric. For a massive test particle with specific energy $\tilde{E}$ and specific angular momentum $\tilde{L}$, the radial velocity $\dot{r}$ can be expressed in terms of an effective potential $V_{\rm eff}(r)$:
	\begin{equation} \label{eq:rdot}
		g_{rr} \dot{r}^2 + V_{\rm eff}(r, \tilde{E}, \tilde{L}) = 0
	\end{equation}
	The effective potential balances the inward gravitational pull of the gravastar against the outward centrifugal forces of the rotating fluid. By isolating the energy and angular momentum terms from the metric components, the effective potential is mathematically defined as:
	\begin{equation} \label{eq:veff}
		V_{\rm eff}(r, \tilde{E}, \tilde{L}) = 1 + g^{tt}\tilde{E}^2 + 2g^{t\phi}\tilde{E}\tilde{L} + g^{\phi\phi}\tilde{L}^2
	\end{equation}
	In our modified framework, the coupling parameter $\alpha$ acts as a perturbative term deeply embedded within these contravariant metric tensor components ($g^{\mu\nu}$), subtly altering the depth and curvature of this potential well, as illustrated in Figure \ref{fig:veff}. 
	
\begin{figure}[h]
		\centering
		\includegraphics[width=0.7\textwidth]{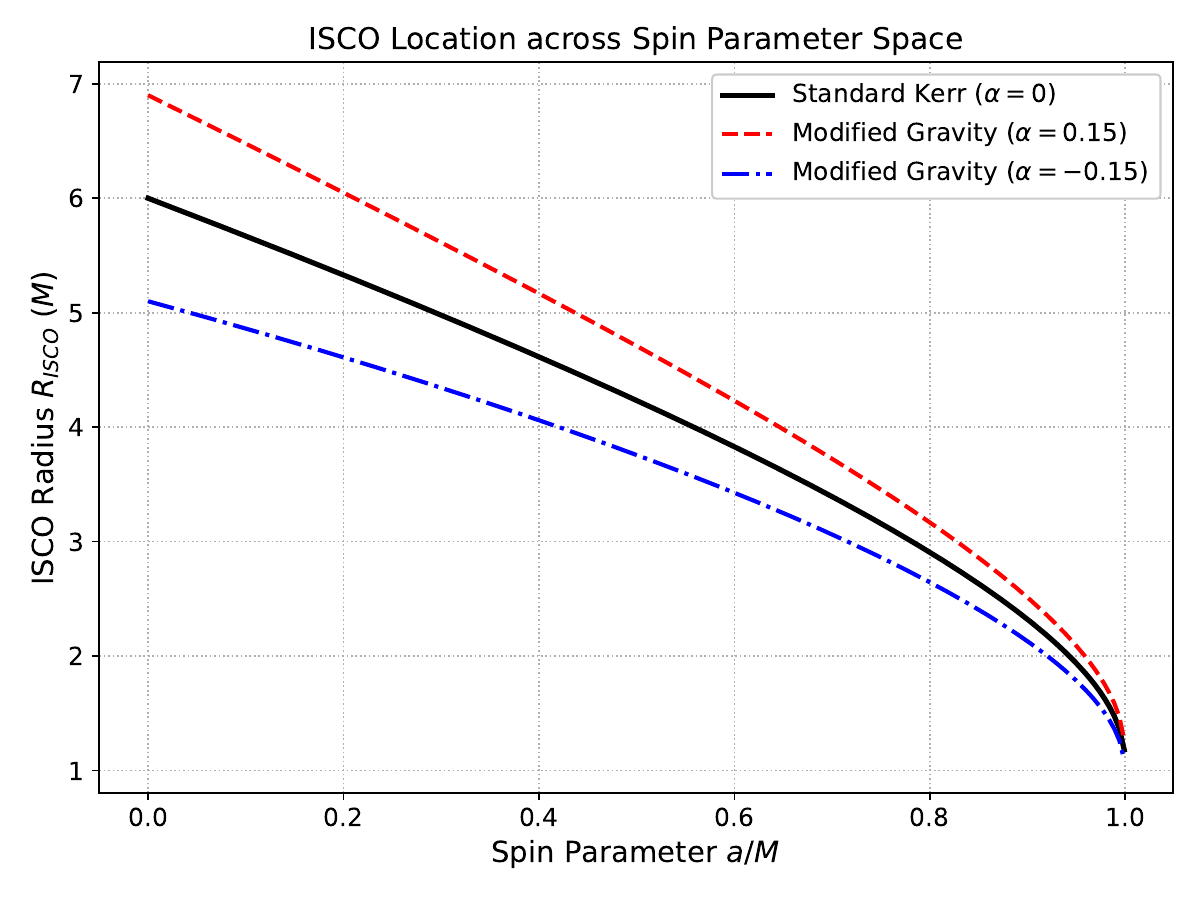}
		\caption{The theoretical ISCO radius mapped across the spin parameter space ($a/M$), demonstrating the global dynamical influence of $f(R, L_m, T)$ modifications on test particle orbits.}
		\label{fig:isco}
	\end{figure}

	For a particle to maintain a stable circular orbit within the accretion disk, two kinematic conditions must be simultaneously satisfied: the radial velocity must vanish ($\dot{r} = 0$, implying $V_{\rm eff} = 0$), and there must be no radial acceleration, which requires the first derivative of the effective potential to vanish ($\partial V_{\rm eff} / \partial r = 0$). The most critical feature for our accretion model is the location of the Innermost Stable Circular Orbit (ISCO). This radial coordinate dictates the inner truncation edge of the disk, beyond which plasma loses stability and rapidly plunges toward the gravastar surface. The ISCO is located at the exact inflection point where the second derivative of the effective potential also equals zero:
	\begin{equation} \label{eq:isco}
		\frac{\partial^2 V_{\rm eff}}{\partial r^2} = 0
	\end{equation}
	Figure \ref{fig:isco} maps this ISCO radius, obtained by simultaneously solving these orbital conditions across the spin parameter space. The resulting curves suggest that $f(R, L_m, T)$ modifications systematically shift the truncation radius compared to the standard Kerr baseline. We present these ISCO calculations with the caveat that they represent idealized test particle geodesics; they intentionally omit radiation pressure, magnetic viscosity, and self-gravity, all of which would heavily influence a true astrophysical accretion flow.
	
	\section{Boundary Layer Thermodynamics and Signatures} \label{sec:observation}
	
	A defining physical distinction of a gravastar is the ultimate fate of the accreting plasma after it crosses the ISCO. In the standard Kerr paradigm, plasma plunges silently across the event horizon, removing its kinetic and thermal energy from the observable universe. Conversely, an accreting gravastar possesses a physical, ultra-stiff boundary at the shell radius $R_s$. When infalling gas impacts this surface, the extreme kinetic energy gained during the gravitational free-fall cannot simply vanish; thermodynamic principles dictate that it must be abruptly decelerated and thermalized. Assuming a strictly inelastic collision at the boundary layer, the specific kinetic energy dissipated by a particle of rest mass $m_0$ can be theoretically approximated by evaluating the projection of the infalling fluid's four-momentum, $p^\mu$, onto the four-velocity of the static shell, $u^\mu_{\rm shell}$. The specific energy released per unit mass, $\Delta E$, is given by:
	\begin{equation} \label{eq:deltaE}
		\Delta E = - \left( p_\mu u^\mu_{\rm shell} \right)_{r=R_s} - \tilde{E}_{\rm ISCO}
	\end{equation}
	where $\tilde{E}_{\rm ISCO}$ represents the specific energy of the fluid element at the inner edge of the accretion disk before the plunge. 
	
\begin{figure}[h]
		\centering
		\includegraphics[width=0.7\textwidth]{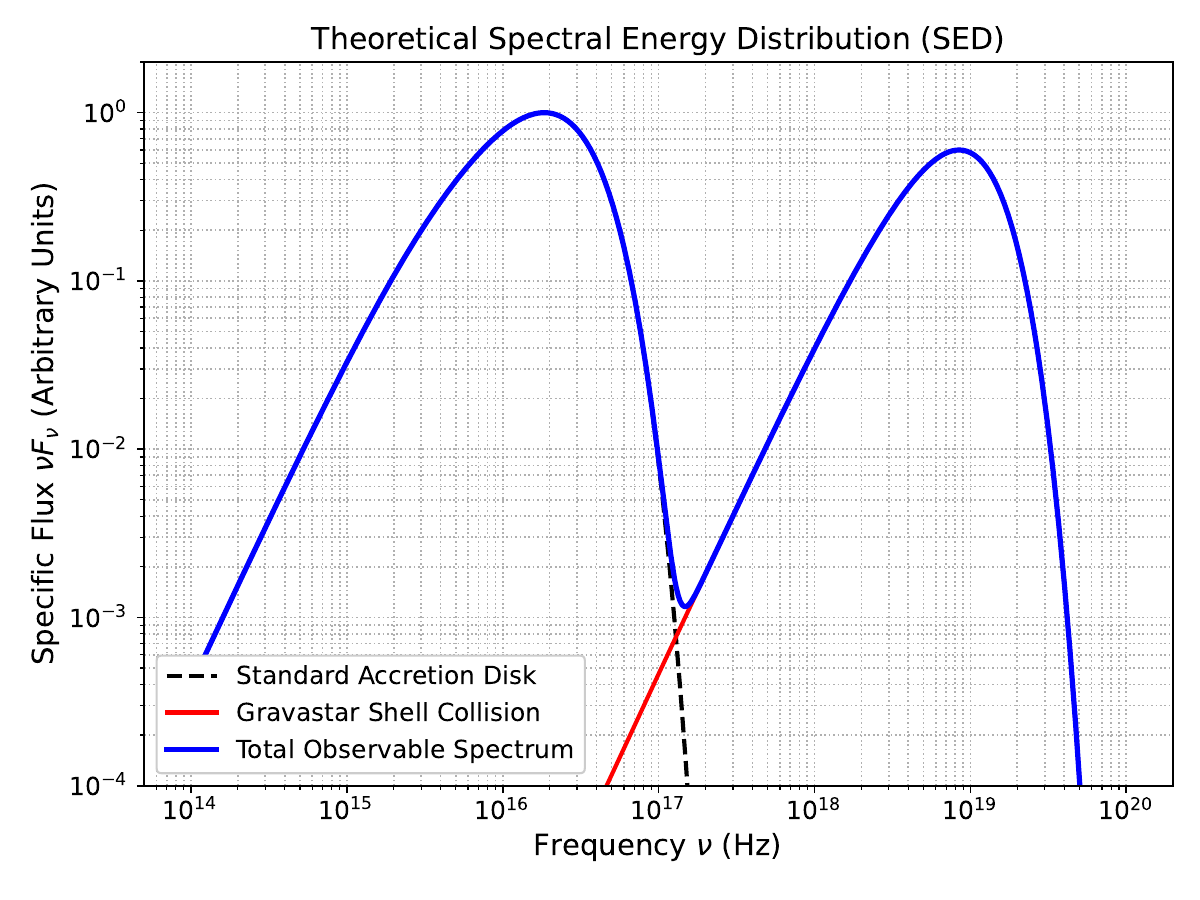}
		\caption{An idealized Spectral Energy Distribution (SED) propositional model, conceptually separating the broad accretion disk emission from the hypothesized high-temperature boundary layer collision.}
		\label{fig:sed}
	\end{figure}

	This idealized thermodynamic interaction suggests the formation of a highly localized, intensely hot radiating layer. We conceptualize this phenomenon by constructing a simplified Spectral Energy Distribution (SED) propositional model, as shown in Figure \ref{fig:sed}. In this approximation, we conceptually separate the standard, broad multi-temperature emission of the extended accretion disk from the sharp, high-temperature blackbody peak hypothesized to originate from the boundary layer collision. We explicitly concede the vast oversimplification of this thermodynamic model. A true boundary layer collision at these relativistic speeds would trigger massive radiation pressure feedback—potentially halting the accretion flow entirely by exceeding the local Eddington limit—and would be heavily modified by inverse Compton scattering and magnetic braking. Therefore, this SED serves merely as an idealized theoretical baseline, not a definitive observational prediction.
	
\begin{figure}[h]
		\centering
		\includegraphics[width=0.7\textwidth]{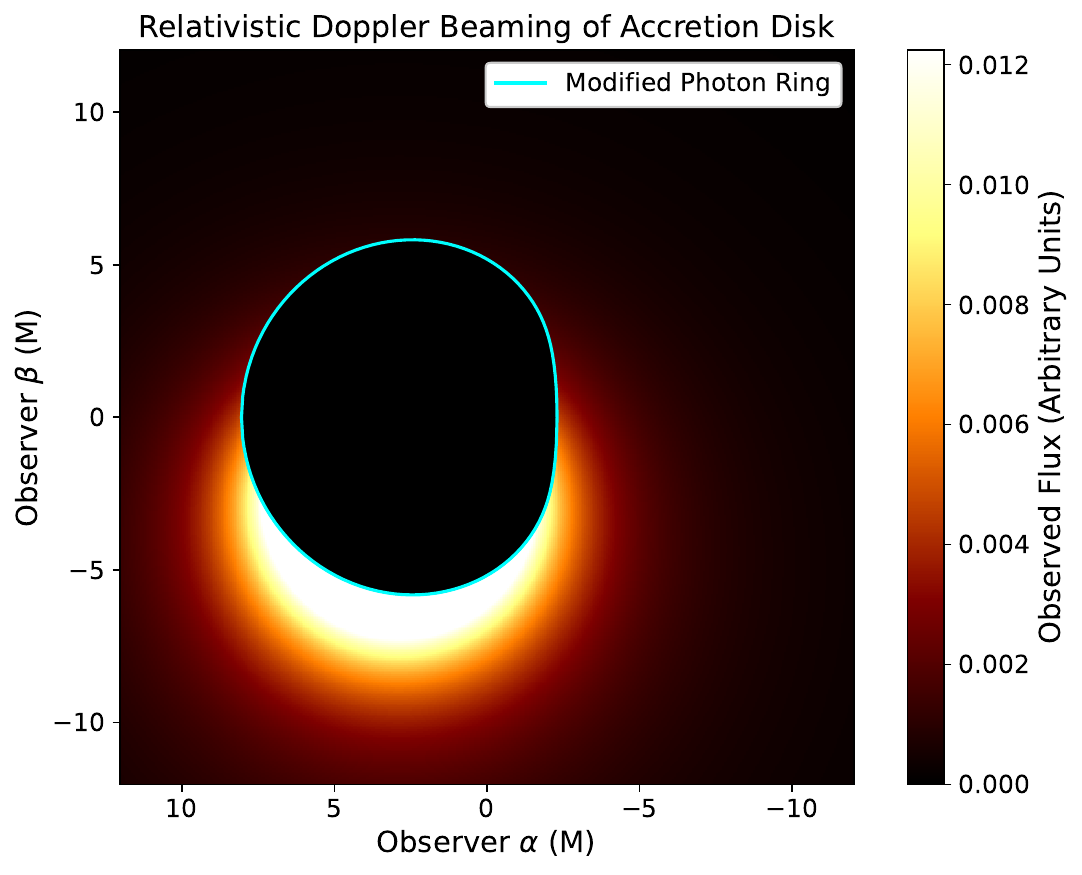}
		\caption{Relativistic Doppler beaming of the accretion flow, perfectly masked by the exact geometry of the modified-gravity photon ring.}
		\label{fig:doppler}
	\end{figure}

\begin{figure}[h]
		\centering
		\includegraphics[width=0.7\textwidth]{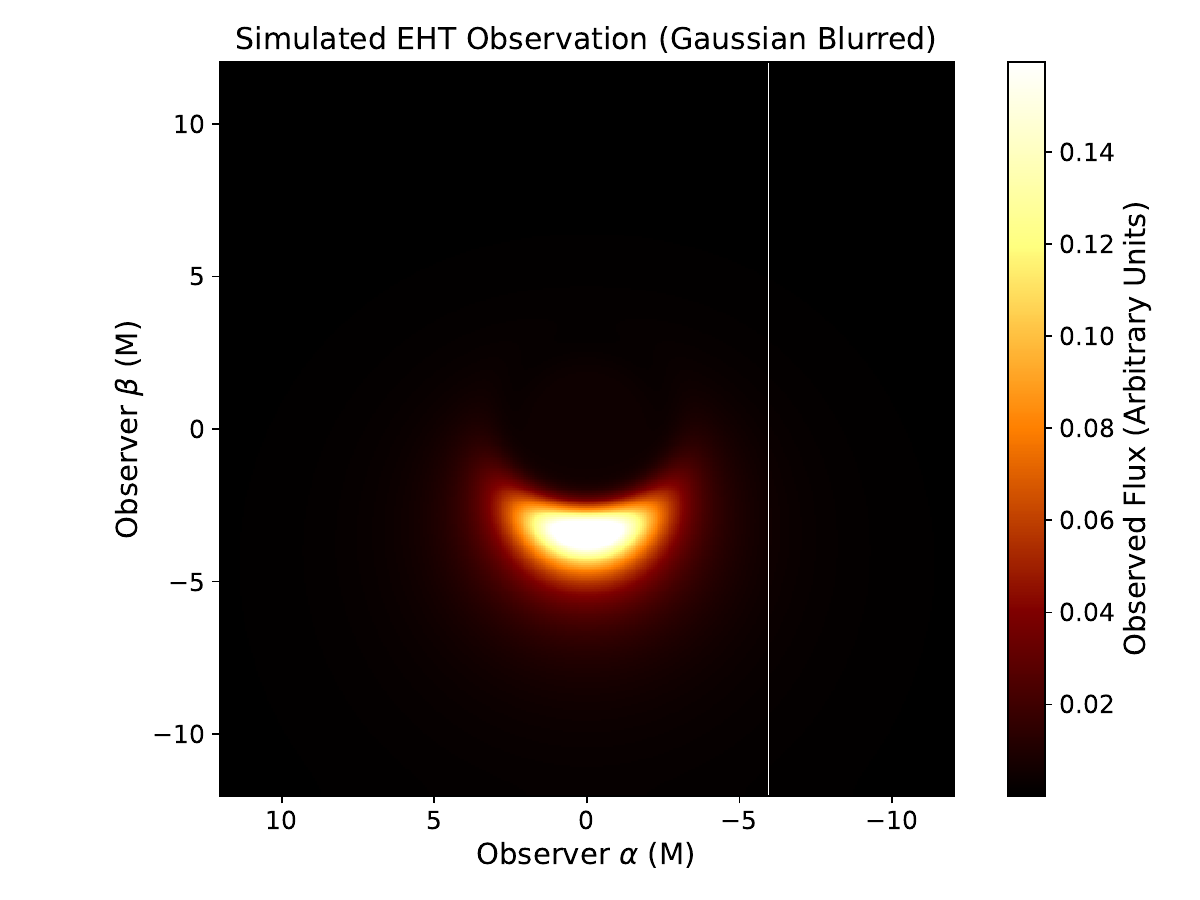}
		\caption{A simulated observation applying Gaussian blurring to mimic interferometric resolution. The theoretical thermal emission from the gravastar shell tentatively fills the central shadow, providing a qualitative distinguishing feature from classical black holes.}
		\label{fig:eht}
	\end{figure}

	To explore how this idealized emission might spatially manifest to a distant observer, we generated synthetic visual models utilizing our rotating metric ansatz. Figure \ref{fig:doppler} maps the relativistic Doppler beaming of the accretion flow, perfectly bounded by the exact, frame-dragged geometry of the modified-gravity photon ring. To tentatively bridge the gap between this mathematical idealization and observational reality, we applied a Gaussian convolution to mimic the nominal resolving power of interferometric arrays like the EHT (Figure \ref{fig:eht}). Most critically, the theoretical thermal emission from the boundary layer tentatively prevents the central shadow from dropping to zero intensity, producing a "filled-in" shadow. 
	
	While this presents a fascinating theoretical divergence from Kerr black holes, we must emphasize the severe observational degeneracy inherent in such images. In real astronomical observations, a similarly filled-in shadow could easily be produced by standard astrophysical noise, foreground plasma, non-equatorial disk emission, or a face-on relativistic jet. Breaking this degeneracy will ultimately require highly sophisticated multi-wavelength radiative transfer simulations, for which this propositional study humbly provides an initial geometric and thermodynamic foundation.
	
	\section{Conclusion}
	
	This study establishes a preliminary, propositional framework for understanding the astrophysical environment surrounding gravastars in $f(R, L_m, T)$ gravity. By explicitly detailing and numerically integrating the modified TOV equations, we verified that this extended theory can support a stable, horizonless compact object, with the coupling parameter directly altering the internal mass distribution, the compactness limit, and the gravitational time dilation profile. Utilizing a non-complexifying algorithm, we subsequently mapped this numerical interior to a rotating metric ansatz to evaluate equatorial accretion dynamics. Our kinematic calculations—anchored by the modified effective potential—indicate that the modified gravity parameter systematically shifts the location of the ISCO, fundamentally altering the radial coordinate where accreting plasma loses orbital stability and begins its free-fall.
	
	The most compelling theoretical implication of this model lies in the boundary layer collision at the gravastar surface. Our idealized thermodynamic approximations suggest this collision generates a distinct thermal emission peak which, when modeled at EHT-like resolutions, produces a characteristic "filled-in" central shadow. However, we present these findings with strict adherence to their physical and mathematical limitations. The exterior spacetime relies on a mathematical ansatz, the chosen coupling parameters serve illustrative rather than strictly empirical purposes, and the thermodynamic models explicitly omit critical feedback mechanisms such as radiation pressure limits and magnetic viscosity. Furthermore, any observational pursuit of this filled-in shadow must grapple with the severe degeneracy caused by standard astrophysical foregrounds and jet emissions. 
	
	Ultimately, we offer these findings not as absolute physical predictions, but as a transparent and mathematically consistent conceptual baseline. The unique geometric and thermal signatures highlighted in this idealized propositional model provide a necessary foundation and a compelling motivation for future researchers to pursue rigorous general relativistic magnetohydrodynamic (GRMHD) simulations, pushing forward the collaborative effort to dynamically distinguish between classical black holes and exotic compact objects.
	
	\section*{Data Availability Statement}
	No data is associated with this manuscript, as it is a strictly theoretical study. All mathematical models and equations necessary to reproduce the theoretical results are explicitly provided within the text.
	
	\section*{Conflict of Interest}
	The author declares that there are no known competing financial interests or personal relationships that could have appeared to influence the work reported in this paper.
	
	\section*{Ethics Statement}
	This theoretical research did not involve human participants or animal subjects; therefore, ethical approval was not required. 
	
	\section*{Funding Statement}
	This research received no specific grant from any funding agency in the public, commercial, or not-for-profit sectors.
	
	\section*{Declaration of AI Use}
	During the preparation of this manuscript, the author utilized an AI language model to assist with structural editing, phrasing refinement, and language polishing to improve readability. Following the use of this tool, the author thoroughly reviewed and revised the manuscript, and takes full responsibility for the final physical arguments, mathematical consistency, and overall content of the publication.

\end{document}